# DocCert: Nostrification, Document Verification and Authenticity Blockchain Solution

Monther Aldwairi
*College of Computer and Information Technology*
*Jordan University of Science and Technology*
Irbid, Jordan
munzer@just.edu.jo

Mohamad Badra
*College of Technological Innovation*
*Zayed University*
Dubai, UAE
mohamd.badra@zu.ac.ae

Rouba Borghol
*Science and Liberal Arts Dept.*
*Rochester Institute of Technology*
Dubai, UAE
rbbcad@rit.edu

*Abstract*— Many institutions and organizations require nostrification and verification of qualification as a prerequisite for hiring. The idea is to recognize the authenticity of a copy or digital document issued by an institution in a foreign country and detect forgeries. Certificates, financial records, health records, official papers and others are often required to be attested from multiple entities in distinct locations. However, in this digital era where most applications happen online, and document copies are uploaded, the traditional signature and seal methods are obsolete. In a matter of minutes and with a simple photo editor, a certificate or document copy may be plagiarized or forged. Blockchain technology offers a decentralized approach to record and verify transactions without the need for huge infrastructure investment. In this paper, we propose a blockchain based nostrification system, where awarding institutions generate a digital certificate, store in a public but permissioned blockchain, where students and other stakeholders may verify. We present a thorough discussion and formal evaluation of the proposed system.

## I. Introduction

Often times, job applicants are required to certify their documents, attest certificates, or equalize a degree or course. The attestation process is lengthy, time consuming, costly and cumbersome, especially when the candidate did graduate long time ago or he graduated from a foreign country, he no longer has access to. Equivalency process requires those attested certificates and transcripts and awards an equivalent and recognized degree. The same process applies for international trade agreements, customs' forms, birth certificates, etc. The process may involve universities, schools, notaries, embassies, departments ministries, education boards, etc. The process takes several months and may be costly if you live abroad. However, after all of that trouble, the final attested or sealed documents may be easily forged digitally. Therefore, not much verification is achieved after this lengthy and costly process [1]. Below is a sample of foreign degree equivalency process requirements.

1. Certified copy of degree (and or all previous degrees) in English or a translated version from an official service. Copies must be attested by
   a. Ministry of education in issuing country,
   b. ministry of exterior in issuing country,
   c. embassy of country seeking equivalency,
   d. ministry of exterior in country of equivalency, and
   e. ministry of education in country of equivalency.
2. Copy of transcript or diploma indicating dates of admission and completion.
3. Copy of passport with visa, entry and exit stamps.
4. Equivalency fees.
5. All original documents.

In most of the above documents, notaries' services may be required. Public and private notaries are authorized by the judicial system to attest documents and certify their originality. Online notaries have been using cameras to verify identity and attest documents. However, the admissibility in court of digital signatures continues to be challenged. Courts often accepts digitally signed documents only when a copy of the originally signed document is presented! [2].

We believe blockchain is a game changer and would present a perfect solution for the online nostrification issue [3]. Blockchain has been made popular with the wide spread of Bitcoin. Bitcoin is one of the earliest and most popular cryptocurrencies. It is a digital currency that can be exchanged between people without the need for a central bank or authority. It benefits from peer-to-peer networks and blockchain technologies to keep an anonymous record of all Bitcoins [4].

Blockchain is a shared immutable ledger for securely recording transaction history. A blockchain could be public or private. As indicated by the name, the blocks are chained, with each block storing one or more transactions. Transactions are kept indefinitely and the blockchain may be queried to verify any transaction, which makes it ideal for nostrification. There have been few attempts to use blockchain for nostrification. EduCTX was one of the first attempts to use blockchain as a higher education credit platform. It is supposed to serve as a centralized repository of all students' records and completed courses [5].

In this paper we propose to use blockchain in a novel manner to implement a secure, shared authenticated and public repository of student records. Students may access their records, so can universities and any other participant who wishes to verify a record. Security and privacy are of the at most importance and therefore access to records is authenticated. The rest of the paper is organized as follows. Section II explains blockchain in more details. Section III surveys the literature, covers the related work and points out advantages and disadvantages of exiting approaches. Section IV discusses the proposed approach and Section V presents the formal evaluation.

## II. BLOCKCHAIN

Blockchain maintains a shared record including full details of every single transaction over a network. Blockchain is based on peer to peer networks, making it distributed and not controlled by any third party [4]. A transaction is any exchange of assets between participants and is represented by a block. Each block tracks and stores data and those blocks are chained together chronologically. Blocks are not editable, which means once a transaction is committed to the blockchain it can no longer be modified. A transaction is reversed by creating a new block, which maintains a timeline of events and changes to the data. Each block contains the transaction data, timestamp, unique hash and the hash of the previous block. The latter maintains the chain and the timestamp ensures timeliness. Unlike databases that are files stored on a single system, blockchain is decentralized and identical copies of the shared ledger are distributed across all participating nodes. This distributed nature of the ledger reduces the chances of data tempering. If a party chooses to add a block to his copy of ledger, it will be inconsistent with all other blockchain participants [6].

Before any block is added to the chain, a consensus of the majority of the endorsing nodes must be reached. Consensus may be through solving a cryptographic puzzle called "proof-of-work", which is the case in many cryptocurrencies. Proof-of-state on the other hand requires validators to hold a cryptocurrency in escrow trusted service. While proof-of-time-elapsed randomizes blocks waiting for trusted environment. Solo-NoOps requires no consensus and validator applies transactions, which may lead to divergent chains or ledgers. Finally, Byzantine-Fault-Tolerance (BFT) achieves consensus in a peer to peer network while some nodes are malicious or faulty [7].

In blockchain for business, we have a shared ledger, where every participant has his own copy. Those ledgers are permissioned and proper credentials are required to access the ledger. The ledger is immutable, in that no participant may tamper with a transaction after it was agreed upon. Transactions cannot be altered deleted or inserted back in time. Smart contracts are a set of business rules in chain code format that when executed a block/transaction is created. The shared ledger has the final say of an asset ownership and provenance. Contrary to cryptocurrencies that emphasize anonymity, blockchain for business is private permissioned network that values identity and permissions over anonymity [8].

Turkanovic et al. explained that higher education institutions (HEI) keep their students' completed courses' records in databases that are structured and only available to institution's staff [4]. Thus, leaving students with limited access where they can only view or print their document. Moreover, these student documents are stored in different standards, which contributes to the problem of transferring student documents to another HEI [9]. Correspondingly, if a student wants to apply for a job in a foreign country, he/she has to translate and nostrificate their academic certificate, which is complex and time consuming. In addition, if a student loses his/her academic certificate, he/she has to visit their HEI and ask for a new copy. Andrejs Rauhvargers discussed qualifications frameworks for recognizing qualification in the European higher education. The paper details the recognition of foreign higher education qualifications [9].

Blockchain is idea for keeping record of student's diploma certificates, transcript, courses, grades, achievements, skills and research experience. All of these may be securely registered in a shared ledger that can be accessed by many institutions or stakeholders. This will help reduce fraud, forgery and false claims [10]. All of the above data can be logged in the form of timely transactions into the shared ledger. The data in the blockchain is permissioned and associated with a student ID, organization ID and stored security in the blockchain. Using blockchain means performance may be sacrificed for secure recordkeeping of transactions. Nonetheless, the blockchain would be much more efficient as opposed to the manual attestation process described earlier [11].

Using blockchain in education might be a new concept, but it surely is very beneficial. It will make it much easier for students to have all of their completed courses certificated, verified and in one place. Not only this will facilitate attestation and verification of qualification but also will help in the cases of credit transfer between institutions. It will be very easy for any workplace anywhere in the world to subscribe to the blockchain and verify graduate credentials, of course with the applicant's consent [12].

## III. RELATED WORK

There are a few research papers concerned with blockchain for nostrification. In this section we summarize each paper and present a critical analysis.

Wibke et al. used blockchain technology to store and handle educational data [13]. They offer the possibility to store different types of immutable educational information on blockchain technology. A total of 58.1 % of the education technologies were based on Ethereum, 3.2 % on Bitcoin, 9.7 % on EOS, and 1.6 % on NEM; 1.6 % used a private blockchain, 4.8 % could be used more than one blockchain and 6.5 % used other blockchain technologies. Their results provide a deeper understanding of blockchain technology in education and serve as a signal to educational stakeholders by underlining the importance of blockchain technology in education.

EduCTX is a blockchain-based higher education credit platform from University of Maribor [5]. This EduCTX platform is anticipated to use ECTX tokens as academic credits. It rests on peer-to-peer networks where the peers of the network are HEI and users of the platform are students and other various organizations. These ECTX tokens represent students credit amount for completed courses. Every student will have an EduCTX wallet for collection of ECTX tokens that will be transferred by his/her HEI. The transferred information is stored in blockchain alongside with the sender's identity with HEI official name, the recipients (which is student and is anonymously presented), the token (course credit value) and the course identification. Therefore, students can access and provide his/her completed courses by directly presenting their blockchain address.

EduCTX is still a prototype based on Ark blockchain platform, and the real-world perception cannot be evaluated. EduCTX enables organizations and students to check academic records of a student's (potential employee) in transparent way. Moreover, since the system is based on blockchain platform, it maintains the possibility of fraud detection and prevention. On the other hand, in the case of a student's losing his/her private key, they have to visit to their home HEI and request a new blockchain address, which is time consuming and almost similar to the current approach for certification. Moreover, it is expected that user and organizations have to protect and backup their private keys, signatures and stamps to be secure, because this platform is yet to have additional level of protection against impersonation.

Gresch et al. from the University of Zurich proposed a blockchain-based Architecture for Transparent Certificate Handing [14]. The work used a questionnaire to shed the light on the wide spread of people with fake diplomas, and how ineffective is the current accreditation system. The system identifies three stake holders: the certificate issuer, companies and institutes wanting to verify diplomas, and the graduates or applications who submitted the diploma. The system has two stages. First, the issuing organization has to create the digital diploma, with one-way hash function and the hash will be stored in a smart contract. Second, the verifier company verifies the authenticity of the document without contacting the university. A prototype was built using an Ethereum blockchain and deployed on University of ZuricH BlockChain (UZHBC). They concluded that granting an organization the ability to issue certificates is one of the most critical aspects of the blockchain. In addition, they only stored the diploma has on the blockchain for privacy concerns. As opposed to storing an encrypted diploma and risking losing the data forever if the key is lost.

Azael Capetllo proposed a blockchain education long-standing model for academic institutions [15]. The paper described the technology of storing student records, which can be shared openly with third parties, offering a safe and lasting record. The technology is strong against data damage or loss, and those third parties can verify student record directly by accessing the University blockchain. Two applications of Blockchain in education have been mentioned in the research paper, the first one is Smart Contracts, and it is to form an autonomous learning experience by consuming an analogy from the financial application of blockchain. The second application is the use of Blockchain to offset the cost learning using peer-to-peer networks, offering financial prize for students offering services to university.

Mike Sharples and John Domingue from University of Nicosia proposed blockchain and Kudos, a distributed system for educational record, reputation and reward [16]. It was the first higher education institution to issue academic certificates whose authenticity can be verified through the Bitcoin blockchain. They proposed to use Bitcoin payments as a reward for academic achievements as tasks such as peer review or assessments. Then they proposed an "educational reputation currency', called Kudos. Each recognized educational institution, innovative organization, and intellectual worker is given an initial award of educational reputation currency, the initial award might be based on some existing metric: Times Higher Education World Reputation Rankings for Universities, H-index for academics, Amazon author rank for published authors etc. An institution could allocate some of its initial fund of Kudos to staff whose reputation it wishes to promote. Each person and institution store its fund of reputation in a virtual wallet on a universal educational blockchain. They used Ethereum smart contracts to implement OpenLearn badges on a private blockchain, where student enroll on courses and institution award them badges.

Wolfgang et al. proposed blockchain in the context of education and certification [17]. The blockchain technology supports counterfeit protection of certificates, easy verification of certificates even if the certification authority no longer exists and automation of monitoring processes for certificates with a time-limited validity. It ensures higher efficiency and improved security for certification authorities through digitization of current processes, issuing and registering of certificates in a blockchain as well as automatic monitoring of certificates. It comprises a blockchain including smart contracts, a public storage holding profile information of certification authorities, a document management system managing the actual payload of certificates tracked by the blockchain and the parties involved in the system, namely accreditation and certification authorities, certifiers, learners and employers.

John Rooksby and Kristiyan Dimitrov from University of Glasgow implemented Ethereum based blockchain technology for permanent and tamper proof grading system [18]. The system was able to store student information on courses enrolled, grades and their final degree. It supported the university specific cryptocurrency called Kelvin Coin. Payment of the cryptocurrency can be made by smart contract to the top performing student in a course. However, there were some drawbacks involved by implementing the system. Scenario-based and focus group evaluation methods were implemented to address the advantages and disadvantage derived from the system. Because universities rely on trust and confidentiality the blockchain system was found to be not trustworthy. Blockchain system was global scope idea, however universities tend to set their own boundaries, at least at institutional level. Moreover, using smart contracts to store grades in blockchain was problematic due the fact that there is no formal algorithm for calculating grades. Unfortunately, the Ethereum Blockchain system needed to change the way of administrative system of the university work. Finally, the prototype of the blockchain system was found to prioritize transparency over efficiency.

Cheng et al. [19] proposed a system that uses Ethereum to generate digital certificates and confirm the eligibility of graduation certificates [19]. The system functions as follows. The HEI enters student's certificate and academic records into the system. The system verifies all the data. The student receives a quick response (QR) code, inquiry number and electronic file of their certificate. Whenever students want to apply for a job or apply for higher education, he/she has only to send the e-certificate alongside with the QR code to the respective organization. The organization can retrieve the student's certificate and academic records once the credentials are verified. Moreover, the QR is used to asses if the certificate is tampered or forged.

There have been several industry projects that were concerned with student records and online digital badges. Many projects capitalized on the opportunity of digital diplomas as countermeasure to fake diplomas. Those projects offered technologies to both mange the complete educational past of students by gathering all digital badges awarded by different academic organizations. Sony Global Education for example, has announced development of a new blockchain for storing academic records [20]. Their platform allows secure sharing of exams results and academic proficiency levels with third-party evaluating organizations. Mozilla Foundation Open Badges are a digital record of the different accomplishments encoded into an image with associated infrastructure for verification [21]. MS Global Learning Consortium was managing this central open source repository of badges with over 1500 participating organizations until 2017. More recently, Mozilla migrated all users to Badgr, as a replacement for Open Badges as the standard verifying credentials [22]. Finally, Acclaim and IBM offer digital badges as a form of organizations recognizing individuals' skills and competencies [23]. Contrary to all of the above industry efforts based in central repositories of badges, BCDiploma is using blockchain to provide security, immutability, ease of use for certifying diplomas and other achievements [24].

All of the above research agreed that counterfeit certificates, credentials and documents is a major problem that can be solved with blockchain. Record all student's academic history from completed courses, skills and qualification in one trusted and secure blockchain is a perfect solution [25]. Yet, all of them tried to tweak current cryptocurrencies blockchain to be used to store certificates and award badges (reputation) and that resulted in low usability. Crypto currencies blockchains and smart contracts are not customized to student records. We propose a permissioned and custom blockchain for business, designed specifically for storing student records or any other document for that matter.

## IV. PROPOSED APPROACH

In this section, we propose an efficient solution that is based on a Merkle tree to provide nostrification and verification of qualifications to guarantee data integrity on the certificates through non-repudiation.

A Merkle hash tree [26] is a data structure used to efficiently verify data integrity and authenticity. As illustrated in Fig. 1, each non-leaf node in the tree, from the bottom up until reaching the root node in the tree, holds the hash of the concatenated hashes of its sub-nodes; example, $h_{12} = h(h_1 | h_2)$. The hash held by the root is represented as the root hash, which can be shared in a trusted way for verification purposes; example $h_{\text{root}} = h(h_{12} | h_{34})$. In [27], a hash calendar is proposed to include the generated root hashes to verify the integrity of the contents of large data structures.

In our proposed solution, we propose forming a Merkle tree where the leaves are documents. The first objective is to provide a periodic publication of the root hash in the blockchain to enhance transparency and protection against any modification in the hashed content, and to provide proof of existence of contents. Each document is certified by its issuer, so we include, in a blockchain, either the hash of that document, or the hash root of a set of documents issued by the same issuer when multiple documents are to be included. The included hash is authenticated by digitally signing it by the same issuer.

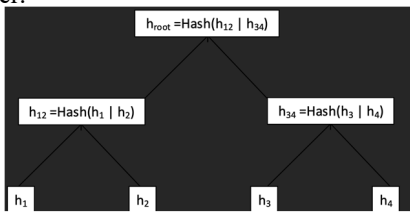

Fig. 1. Example of Merkle tree with 4 leaves (depth = 2).

Interested parties can authenticate the existence of any document and the verification is legally acceptable. The verification process relies on the authentication path of a given node in the tree to validate the hashed content held by the node, without Merkle tree traversal [28]. The authentication path of a node in the tree consists of a set of siblings on the path from that node to the root. The content of a document can be authenticated using the hashed contents held by the root node and by the corresponding authentication path as well. For example, and with reference to Fig.1, to verify $h_1$, the verifier needs the values of $h_{\text{root}}$, and the authentication path $h_{34}$ and $h_2$. Hence, the verifier computes $h'_{12} = h(h'_1 | h_2)$ and $h'_{\text{root}} = h(h'_{12} | h_{34})$, and then compares $h'_{\text{root}}$ with $h_{\text{root}}$ for equality. Choosing the hash function or algorithm [29] depends on many factors such as speed, digest length, number of rounds, collisions and ease of implementation both in hardware and software [30].

*A. Transaction Structure*

When a transaction is generated by our entities, it should include the hash root and a set of hash values, where each hash is the digest of a document belonging to the user.

$h_{\text{root}}$   Set of hash values (i.e., $h_1$, $h_2$, …)

*B. Nostrification's Generation of Document's Qualification*

The proposed system is very versatile and can be applied to any document and authentication process. In this section, we describe our solution using three different scenarios. In the first one, we describe the case where the user has several documents issued by the same entities, whereas in the second case, the user has several documents issued by different entities. The third scenario is concerned with the case of one document that will be certified by a hierarchy of different institutions.

The proposed nostrification solutions supports 3 cases or operating scenarios, because of space limitation we present cases 1 and 3.

*Case 1*

In this case, the issuer (entity) of the documents will form a Merkle tree where each leaf is the hash of a document (Fig. 2). Next, the entity will generate the authentication path for each intermediate node in the tree, the digest of each document, and the hash root of the tree. Then the entity will sign the hash root and publish it along with the hash value of each document in the blockchain. The entity will next issue the documents to the user after stamping each document. The stamp consists of adding to each document, the identifier of the transaction that is already stored in the blockchain.

*Case 3*

This case is similar to the two previous cases; however, each tree is dedicated to one document only and each layer of the tree is associated to an organization that will certify the document. Each organization has a private and a public key. When we want to nostrify a document, we start by selecting all organizations that will certify the document. Then, with the hash of the document, we sign that hash with the private key of the first organization. Finally, we create the right node of the layer with the pair constituted by the signature and the location of the public key of the organization, required to verify the signature. The parent node of the layer is created by hashing the result of the concatenation of the hash from the left node and the signature from the right node. The process is then repeated for each organization. When a layer has been created for each organization, we calculate the last hash that will become the hash root and we have the Merkle tree (Fig. 3).

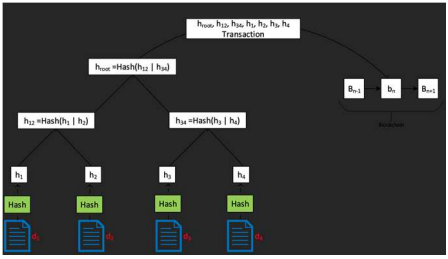

Fig. 2. Nostrification of several documents issued by the same entities.

As everyone is able to get public keys of organization with the location included in the tree, it is very simple to verify the authenticity of a document certify by any number of organizations.

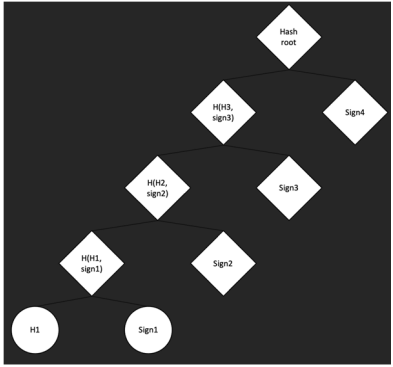

Fig. 3. Nostrification of a single document by different entities.

## C. Nostrification's Verification of Document's Qualification

Any third party who is willing to verify any document that is shown by the user, the third party shall first query the blockchain to extract the root hash and the set of hash values as well, using the transaction identifier stored on the document presented by the user. Then, the third party generates the digest of the presented document and compares it for equality with the one of the hash values stored in the extracted set of the hash values. Then, the third party regenerates the hash root and compares the result for equality with the hash root that is downloaded from blockchain. Then, it verifies the signature on the hash root that is generated by the document issuer, and if the signature is validate, the third party approve the document.

In case a third party is willing to verify more than one document that were nostrified by more than one entity, then the user should send the transaction identifier of the most recent nostrified document, which includes a hash root, and a set of hash values. This latter set includes the hash of the most recent nostrified document and the hash of any other document belonging to the user and nostrified prior to nostrifying the most recent nostrified document.

## V. IMPLEMENTATION AND SECURITY ANALYSIS

In this section, we present a detailed analysis of the proposed approach's security and we demonstrate its effectiveness in providing in providing long-term integrity protection, proof of existence, authenticity, non-repudiation and privacy. Additionally, we evaluate the efficiency in terms of the processor and time usage.

We start with one of the most popular attacks on data integrity, False Data Injection (FDI). In FDI the attackers intentionally change the data in such a way that the receiver will be unable to detect forged data. Blockchain by its design is secured against tampering and revision, which makes it very difficult to the adversary to inject or add forged or malformed document into the blockchain. In addition, the signature of the issuer over the document, make it much more difficult, even impossible, to inject malformed data into the blockchain.

In addition to the long-term data integrity and the proof of existence, our solution ensures the authenticity and the non-repudiation of origin because the hash root will be signed by the last document's issuer when we have several documents from different entities, and by the documents' issuer when those documents were issued by the same issuer. It is worth noting that the issued documents will be always valid if the issuer's certificate will expire or revoked. In fact, our solution leverages blockchain approach properties to provide long-term integrity of documents.

Privacy concern usually arises in many applications, particularly in is a public distributed database like the blockchain. The privacy concerns are mostly related to the publication of the user's documents. In our solution, the privacy is preserved since the digest of the documents are stored in the blockchain, but not the document itself. Hence, the adversary needs to crack the digest in order to find out the original document.

Our approach maintains the above security services while reducing the computation overhead. In fact, instead of generating a signature for each document issued by the same entity to the user, the entity will only need to sign the hash root. However, our solution will introduce very limited computation overhead related to the hash function that will be applied to generate the hash of each document that is required to compute the hash root. But consider the asymmetric encryption computational overhead when compared to the hash function computational overhead, this latter is usually negligible.

The system proof of concept was implemented using Python v3, Flask webserver, and Ganache is used to create the blockchain test server. To evaluate the efficiency of the system we measure the CPU usage, memory consumption, and time for adding a document and the nostrification process. The *PyCharm* IDE was used for measurements and average of five runs. In Table I, we can observe that adding a user will relatively take more time because of the deployment of the contract on the blockchain.

Our system proof of concept allows everyone to verify the authenticity of a document after accessing both the authentication path and the hash root, which are stored into a smart contract that is publicly available on a blockchain. Our used smart contracts are based on the same cryptography technology being used by cryptocurrency; therefore, they have the same level of security. All details of deployments or updates of contracts are written into transactions to help finding the data at any time. Particularly, it allows storing data like the username and the user's Merkle Tree. When the issuing institution adds a user along with its documents to the blockchain, a contract is deployed, and a transaction initializes it with the data. At any time, the issuing institution can add several documents to the user profile, in which the user's Merkle Tree is then updated, and a new transaction is also needed to update the data stored on the blockchain.

TABLE I. SIMULATED PERFORMANCE

|  | CPU | Memory (MB) | Add Time (s) 1 user & 4 documents | Time (s) Nostrification |
|---|---|---|---|---|
| Case1 | 2% | 37 | 7.5 | 0.001 |
| Case3 | 5% | 49 | 0.22 | 0.028 |

## VI. CONCLUSIONS

In this paper, we proposed a blockchain-based nostrification system, where awarding institutions generate a digital certificate, store in a public but permissioned blockchain, where students and other stakeholders may verify. We present a thorough discussion and formal evaluation of the proposed system. In addition, we implemented a prototype of the solutions supporting 3 use-cases. The formal analysis shows resistance to all sorts of common attacks with excellent performance in terms of CPU and memory usage as well as negligible blockchain programing and query times.

ACKNOWLEDGMENT

This project was supported in part by Zayed University Research incentive grant #R22018.